\documentclass{iau_FM}
\usepackage{amsmath}
\usepackage{graphicx}
\usepackage{longtable}
\usepackage{multirow}
\usepackage{natbib}
\usepackage{journals}
\usepackage{hyperref} 
\usepackage{blindtext}
\usepackage{titlesec}
\usepackage{comment}

\title[chapter:hardware] {Real-time RFI Mitigation Techniques in Radio Astronomy and their Practical Limitations}
\author[Buch, Gunaratne, Hellbourg, Viou and Winkel]   
{
 Kaushal D. Buch$^1$
 \and Thushara Gunaratne$^2$
 \and Gregory Hellbourg$^3$
 \and Cedric Viou$^4$
 \and Benjamin Winkel$^5$
}

\affiliation{
$^1$ Giant Metrewave Radio Telescope, NCRA-TIFR, Pune, India \\ email: {\tt kdbuch@gmrt.ncra.tifr.res.in}\\[\affilskip]
$^2$ Herzberg Astronomy and Astrophysics Research Center, National Research Council Canada \\ email: {\tt Thushara.Gunaratne@nrc-cnrc.gc.ca} \\[\affilskip]
$^3$ California Institute of Technology, Pasadena, CA, USA \\ email: {\tt ghellbourg@astro.caltech.edu}\\[\affilskip]
$^4$ Observatoire Radioastronomique de Nançay (ORN), Observatoire de Paris, Université PSL, Université d'Orléans, CNRS, Nançay, France \\ email: {\tt Cedric.Viou@obs-nancay.fr} \\[\affilskip]
$^5$ Max-Planck-Institut für Radioastronomie - Effelsberg radio observatory \\ email: {\tt bwinkel@mpifr-bonn.mpg.de} \\[\affilskip]
}
\begin{document}

\maketitle

\begin{abstract}
Radio astronomy is facing critical challenges due to an ever-increasing human-made signal density filling up the radio spectrum. With the rise of satellites, mobile networks, and other wireless technologies, radio telescopes are struggling with radio frequency interference (RFI), which can masquerade, block or distort astronomical signals. In this chapter, we explain where RFI comes from, how it affects observations, and discuss different ways to reduce or remove interference. The techniques presented here reflect the state of the art in real-time RFI mitigation at the time of publication and include methods such as filtering, digital processing, and optimal scheduling. The proposed catalogue also explores new ideas like satellite avoidance through scheduling, the use of intelligent surfaces to block interference, and advanced computer algorithms to clean up data. The chapter also highlights the need for strong cooperation between astronomers and spectrum regulators to protect radio frequencies for future discoveries. By combining technical solutions and better policies, we can help ensure that radio astronomy continues to provide important insights into the universe.
\end{abstract}


\section{Introduction}
\label{section:hardware: introduction}


\subsection{Basics of RFI in radio astronomy}
The radiometer equation is fundamental in radio astronomy for determining the sensitivity of a radio telescope. It quantifies the sensitivity of a radio telescope as a function of the receiver system temperature, bandwidth, and observation time, offering insight into the minimum detectable signal. The equation is given by:
\[ \Delta S = \frac{2 k_B T_{\text{sys}}}{A_{\text{eff}} \sqrt{2 \Delta f \tau}} \]
where \( \Delta S \) is the minimum detectable flux density in Jansky (1 Jy = $10^{-26}$Wm$^2$Hz$^{-1}$), \( k_B =1.38 \times 10^{-23} \;\text{J} / \text{K}\) is the Boltzmann constant, \( T_{\text{sys}} \) is the system temperature in Kelvin (K), \( A_{\text{eff}} \) is the effective area of the telescope in $m^2$, \( \Delta f \) is the observed bandwidth in Hz, and \( \tau \) is the integration time in seconds (s). This equation highlights the importance of maintaining a low system temperature, utilising a large effective area, operating over a broad bandwidth, and employing a long integration time to enhance the sensitivity and accuracy of astronomical observations.

In the context of interferometry, where signals are correlated between pairs of antennas rather than detected by a single dish, the sensitivity of a single visibility measurement is given by the interferometric radiometer equation:

\[ \Delta S = \frac{2 k_B T_{\text{sys}}}{\eta_s A_{\text{eff}} \sqrt{ \Delta f \, \tau }} \]
Here, $\eta_s$ is the system efficiency factor, typically close to unity. The absence of the $\sqrt{2}$ factor (compared to the single-dish case) reflects that real and imaginary components of visibilities are treated separately. In an array with $N$ antennas, the overall image sensitivity improves with the number of independent baselines, scaling approximately as $\sqrt{N(N-1)}$ for a complete array.

These two forms of the radiometer equation form the theoretical foundation for quantifying sensitivity and for designing optimal observational strategies in radio astronomy. The sensitivity that modern radio astronomy systems can achieve is unrivalled in other radio services and applications. However, this fantastic achievement also comes at a cost. Observations are highly susceptible to various human-made transmissions and other forms of electromagnetic radiation.
For instance, the downlink of a single low-Earth orbit satellite ($\approx$ 550 km altitude) in the cellular band, concentrated into a 10 MHz channel, can appear over 50 dB brighter than the quiet Sun ($\approx$ 0.75 MJy at 1 GHz) and more than 80 dB brighter than Cassiopeia A ($\approx$ 2.4 kJy at 1 GHz), which are respectively the brightest and second-brightest natural radio sources in the sky.

The sensitivity requirements of modern radio astronomy compel astronomers to explore frequencies well beyond the protected passive bands owing to the scarcity, limited bandwidth, and incomplete coverage of these designated ranges. Astronomers usually refer to the anthropogenic signals found beyond these protected bands as RFI. Still, it has to be pointed out that in legal terms, most of the disturbing features in the radio telescope data are not considered as RFI by administrations and other spectrum management stakeholders, as the operators of radio transmitters usually have a license, and the wanted transmissions, e.g., cell phone carriers, are perfectly legal. In spectrum management, RFI specifically means any kind of unwanted emission, such as near-spectral sidelobes, harmonics or intermodulation products, which can leak into the allocated band of another radio service.

While the Radiocommunication Sector of the International Telecommunication Union (ITU-R) has acknowledged the requirements of radio astronomy decades ago, the total amount of spectrum that is allocated to the Radio Astronomy Service (RAS) is low. As a consequence, in addition to measurements in the RAS bands, we are opportunistic -- observing the parts of the spectrum outside of the protected bands wherever and whenever possible at an observatory, as not every active radio service is equally utilised everywhere. 

With more and more digital transmissions, however, the efficiency of spectrum usage has increased significantly in recent years. Therefore, the need increases to deal with human-made features in our datasets, which can range from flagging affected data samples to attempting to fully remove them.

\begin{figure}
    \centering
    \includegraphics[scale=0.25]{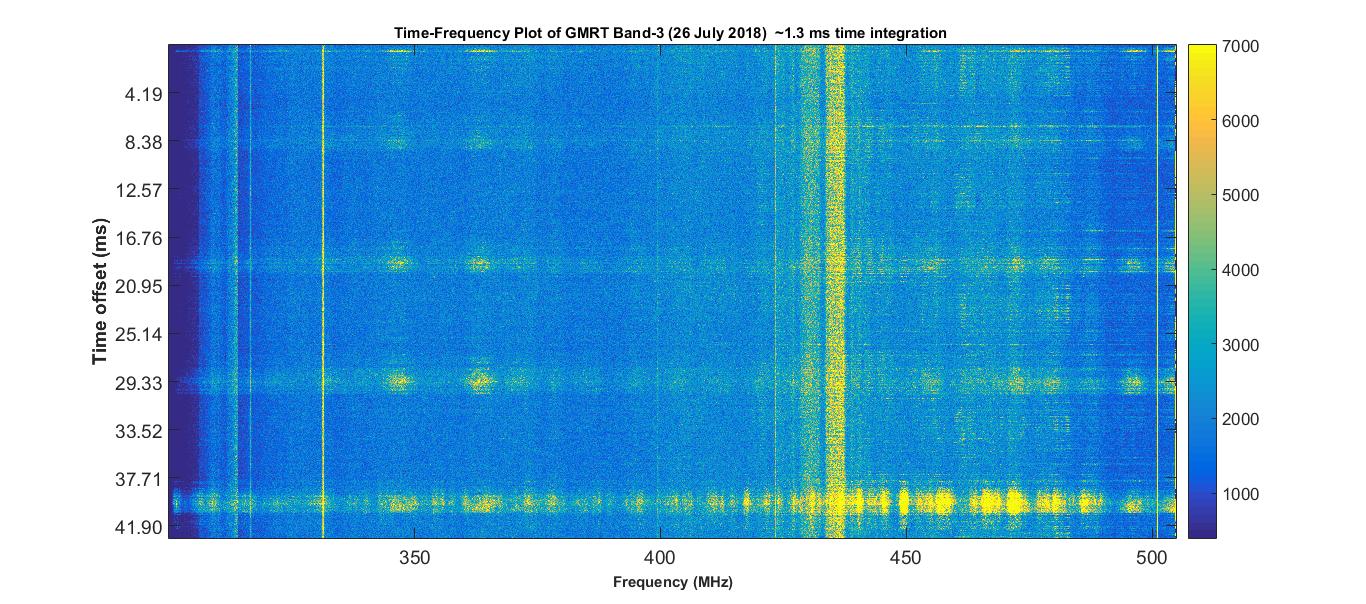}
    \caption{Spectrogram of upgraded Giant Metrewave Radio Telescope (uGMRT) \citep{gupta2017upgraded} antenna showing broadband (powerline RFI) and narrowband (communication transmitters) RFI in the 250-500 MHz band. Broadband powerline RFI is seen to be repeated every 10ms (submultiple of 50Hz powerline frequency). Note that the astronomical signal is buried below the noise floor and cannot be identified in this plot.}
    \label{fig:ugmrt-rfi}
\end{figure}

Generally, in radio astronomy, RFI is classified based on spectral occupancy. The primary classification is broadband and narrowband. An example of broadband and narrowband RFI is shown in Figure~\ref{fig:ugmrt-rfi}.

It should be noted that the best form of interference mitigation is still to avoid them in the first place. Radio astronomers have done their part by constructing telescopes in the most remote locations physically accessible, with as little radio noise background as possible. Often, site searches entailed dedicated spectrum monitoring to find the quietest places. Making use of terrain shielding provides additional protection. Furthermore, astronomers participate in national and international spectrum management forums such as the International Telecommunication Union (ITU) to advocate for protection from new applications in the bands allocated to the RAS.

Only if all these efforts fail -- and this is mostly the case -- actual RFI mitigation needs to be considered.


\subsection{Levels of RFI}
\label{subsection:hardware:introduction: levels}
A radio astronomy receiver chain begins with a feed that collects celestial radio waves, directing them to a low-noise amplifier (LNA). The LNA amplifies these weak signals with minimal added noise, ensuring data integrity. After amplification, the signal passes through a bandpass filter to isolate the frequency of interest, followed by downconversion, when applicable, to an intermediate frequency (IF) using a mixer and local oscillator. The IF or baseband signal is further amplified, filtered, and digitized by an analog-to-digital converter (ADC), though modern systems increasingly digitize directly at the radio frequency (RF) to minimize signal degradation.

RFI impacts the receiver chain at varying levels:
\begin{itemize}
\item \emph{Level 0}: RFI that appears undetectable relative to the system sensitivity, for some observations. While such signals may fall below the detection threshold in a single observation, they can still introduce subtle biases, particularly in statistical or detection-limited experiments near the noise floor, through persistent systematic effects. According to the radio interferometer measurement equation, even low-level signals can bias results if they correlate with the observing process. Thus, \emph{Level 0} should be understood as RFI whose power spectral density is at least an order of magnitude below the sensitivity threshold of the instrument. It is often referred to as “unaffecting RFI,” though its actual impact may only be assessable through long integrations or modelling.
\item \emph{Level 1}: Detectable RFI adds unwanted contributions, requiring excision to preserve data integrity, but this can reduce sensitivity and hinder the detection of sparse astronomical signals or calibration accuracy.
\item \emph{Level 2}: Stronger RFI pushes the LNA into non-linear operation, generating harmonics and intermodulation products. Mitigation involves excising data and using attenuators, further degrading sensitivity.
\item \emph{Level 3}: Severe RFI causes ADC saturation, resulting in irrecoverable signal artefacts. Mitigation typically includes attenuators at the cost of a loss in sensitivity or ADC direct sampling with a high dynamic range.
\item \emph{Level 4}: Extreme RFI physically damages receiver components, requiring the telescope to cease operation.
\end{itemize}
Spectrum management aims to ensure RFI remains at or below Level 0 within primary radio astronomy bands as defined by international allocations, as well as within nationally or voluntarily protected zones such as Radio Quiet Zones (RQZs). In these cases, signals may fall below the threshold to be considered interference at all. Outside of these protected bands or regions, all levels of RFI may occur depending on the local spectral environment. This highlights the persistent challenge of protecting sensitive radio astronomy observations from both in-band and out-of-band emissions.

\subsection{Motivation for real-time RFI mitigation}
\label{subsection:hardware:introduction: motivations}

\begin{figure}
    \centering
    \includegraphics[height=.20\textheight]{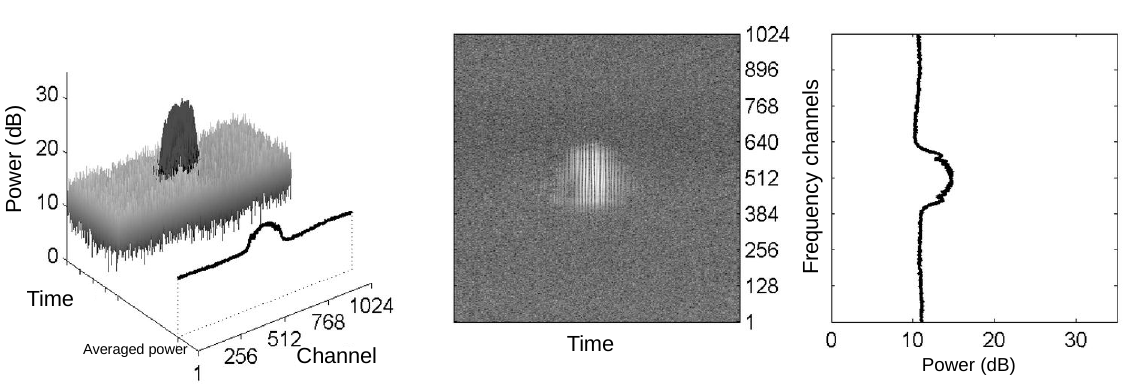}
    \caption{One of the 5-s periodic bursts received from a radar system.  The burst is a group of many periodic short impulses, sparse in time, but broad.  }
    \label{fig:rfi_example_radar}
\end{figure}

RFI mitigation in a radio observatory is carried out in different ways \citep{ford2014rfi}. Real-time RFI mitigation usually happens in the analog domain or at the highest time resolution, i.e. close to the Nyquist rate, which helps minimize the data corruption due to sparse RFI in downstream signal processing with minimal loss of astronomical data, thereby improving and enhancing the quality and accuracy of astronomical measurements. Other benefits from real-time RFI mitigation are as follows:

\begin{enumerate}
\item For time-domain impulsive RFI, the energy spreads across the observing band in the frequency domain, making it impossible to mitigate in the spectral or post-correlation domain.  See Figure~\ref{fig:rfi_example_radar} for an illustration of a radar system.

\item Correlated RFI is best treated in the pre-correlation domain to reduce its ill effects on the astronomical data.

\item RFI is mostly non-random; hence, its early removal helps follow the radiometer equation, which is crucial to achieving the desired sensitivity of the telescope.

\item It helps facilitate adjustments to the telescope signal processing chain in response to a dynamic RFI environment. This adaptability is crucial in maintaining the continuity of observations and ensuring that data collection is optimized even in transient or sporadic RFI.
\end{enumerate}

\begin{figure}
    \centering
    \includegraphics[scale=0.8]{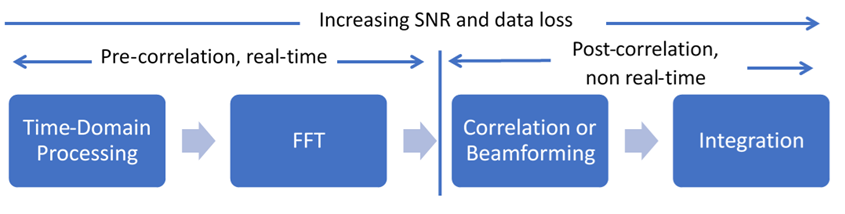}
    \caption{SNR versus data loss in a typical telescope receiver system}
    \label{fig:real-time-rfi}
\end{figure}

Figure~\ref{fig:real-time-rfi} illustrates the increase in data loss in a typical radio telescope receiver signal processing chain.

Real-time RFI mitigation strategies fall into three broad categories: \textit{excision}, \textit{subtraction}, and \textit{coordination}. Excision techniques identify and remove contaminated samples before averaging, reducing downstream corruption but resulting in irreversible data loss and potential statistical biases \citep{hugo2022tricolouroptimizedsumthresholdflagger,10464448}. Subtraction methods attempt to preserve the astronomical signal by estimating and removing the interference, yet are limited by residuals due to imperfect modelling and front-end nonlinearities \citep{ellingson2022coherent,chakraborty2024low}. Coordination approaches mitigate RFI proactively—e.g., via predictive scheduling or emitter cooperation—but may constrain telescope operations \citep{hellbourg2024assessing}. These trade-offs are summarized in Table~\ref{real-time-tech}.

\section{Catalog of real-time mitigation techniques}
\label{section:hardware:catalog}

In this chapter, we provide a comprehensive catalogue of state-of-the-art real-time RFI mitigation techniques. For this discussion, real-time is defined as techniques applied directly to incoming data streams during acquisition and that cannot be reversed or ``unapplied'' at a later stage. Unlike offline or post-processing methods, where raw data remain intact for subsequent re-analysis, real-time techniques irrevocably transform the recorded data. Consequently, the effectiveness and precision of these methods directly determine the scientific or operational value of the collected datasets.

\subsection{Existing real-time implementations}
\label{subsection:hardware:catalog:existing}
\subsubsection{Static and dynamic scheduling and avoidance}
Effective scheduling is pivotal in mitigating the impact of RFI on radio telescope observations.
Many modern observatories implement some form of scheduling system to align astronomical targets with telescope configurations, which can help optimize observation strategies. When such systems incorporate knowledge of known fixed sources of RFI, such as terrestrial emitters and geostationary satellites, they can strategically plan observations to minimize interference.
Fixed sources are well-documented, and their positions can be accounted for during the scheduling process. This is achieved by aligning the telescope's sequence of sky pointings with its directivity pattern, ensuring that these known interference sources are avoided during observations.

In addition to accounting for fixed sources, many telescopes are increasingly adopting dynamic scheduling strategies to address moving sources of RFI (e.g. \cite{guzman2016status,luo2018cost}), such as medium Earth orbit (MEO) satellites and aircraft. These strategies leverage publicly available satellite position data and aircraft tracking information from ADS-B transmissions to dynamically adjust telescope operations. However, avoiding low Earth orbit (LEO) satellites presents a significant challenge due to their high angular velocities and vast numbers. Dedicated solutions for mitigating the impact of LEO satellites, which are under active investigation, may provide more robust approaches in the future (see \ref{subsubsec:avoidance}).

For spatially confined sources of interference, such as satellites that affect the Upgraded Giant Metrewave Radio Telescope (uGMRT) in India during specific intervals when the telescope beam intersects with satellite beams, a real-time satellite monitoring system has been implemented. This system provides advanced notifications about satellite passes, their typical duration and logs relevant data. Additionally, it raises alarms in the control room to alert operators of potential interference \citep{raybole2016real}. By integrating this system into the operational workflow, the uGMRT can proactively reduce the impact of satellite RFI, minimizing disruptions to scientific observations. However, it is important to note that such systems primarily serve as early-warning tools. They do not typically trigger automatic interruption of ongoing observations. For example, the uGMRT does not repoint the telescope mid-integration, and RFI occurring during long-duration observations may still degrade data quality.

Dynamic interference avoidance strategies are gaining traction, particularly for next-generation telescopes. Facilities such as the Five-hundred-meter Aperture Spherical Telescope (FAST) in China, the Next Generation Very Large Array (ngVLA) in the United States, and the Deep Synoptic Array (DSA-2000) also in the United States are actively exploring these approaches. Similarly, the Australian Square Kilometre Array Pathfinder (ASKAP) is developing an RFI avoidance scheme specifically for addressing atmospheric ducting events, which can propagate RFI over vast distances.

Dynamic scheduling has also been implemented in operational telescopes like the Green Bank Telescope (GBT) in the United States and the MeerKAT telescope in South Africa. These telescopes utilize real-time spectrum monitoring to adapt their observation schedules based on the prevailing RFI conditions. For example, the GBT adjusts its schedule to avoid heavily contaminated frequency bands, allowing observations to continue in cleaner spectral windows. Similarly, MeerKAT shifts its focus during periods of high RFI, such as interference caused by local broadcasting or transient sources like aircraft. This adaptive approach ensures optimal use of telescope time by prioritizing frequencies least affected by interference.

Dynamic scheduling frameworks not only help in protecting observations from RFI but also enhance the overall quality of scientific data. By continuously monitoring the spectrum and adjusting observation plans in real time, telescopes maximize data integrity and improve sensitivity to faint cosmic signals. These strategies are particularly valuable in environments where RFI fluctuates frequently, ensuring that telescopes can maintain high observational efficiency and reliability across diverse conditions.

\subsubsection{Adaptive analog attenuators}

The Owens Valley Radio Observatory Long Wavelength Array, in the United States, employs adaptive attenuators to maintain stable baseline levels and prevent saturation in response to strong citizen broadcast RFI, which skips through the atmosphere during the day but not at night. Adaptive attenuators play a critical role in ensuring that the telescope array can handle this variable interference environment by adapting the signal attenuation during daytime and nighttime. This approach effectively
prevents saturation and preserves operational continuity. However, it reduces sensitivity and dynamic range, as strong attenuation can limit the number of bits available for encoding weak astronomical signals in the ADC. This may, in turn, introduce quantization artefacts or reduce measurement fidelity, especially for faint sources. As a result, deep, sensitive observations are typically restricted to nighttime hours when the RFI is less intense.
However, observations of particularly strong sources, such as the Sun, remain unaffected by this daytime mitigation strategy, allowing these studies to proceed with minimal interference.

\subsubsection{Front-end notch filters}

Notch filters are a widely used technique for mitigating strong narrowband RFI in radio astronomy, addressing the challenges posed by interference from various terrestrial and satellite-based sources.
However, they are generally considered a last-resort solution, particularly when placed before the first stage of amplification, due to their significant impact on receiver sensitivity.
Because of their self-generated noise and insertion loss, such filters can increase the system temperature by several kelvin, degrading overall performance.
For example, introducing a notch filter with an insertion loss of 0.5-1 dB can raise the receiver temperature by 10-20 K, a non-negligible penalty in sensitive astronomical receivers.

Despite these drawbacks, notch filters play a crucial role in preserving the dynamic range and linearity of receiver systems by attenuating specific interfering frequencies when positioned at critical stages in the signal chain. 

At the uGMRT, band-reject (notch) filters are incorporated into the front end of the receiver system. These filters are designed to mitigate strong and persistent RFI from sources such as short-range transmissions, mobile communications, and broadcast television, ensuring the receiver operates within its linear range. Ongoing efforts aim to position these filters before the Low Noise Amplifier (LNA) to enhance their effectiveness by preventing saturation from strong interfering signals \citep{sureshkumar2016rfi}. However, this approach is only considered for extreme cases where the power of the RFI would otherwise lead to severe non-linearities or hardware damage, given the sensitivity loss involved.

Similarly, the GBT employs switchable notch filters to target and suppress specific frequency bands prone to RFI contamination. For instance, within its C-band (4–8 GHz) receiver system, a notch filter is strategically placed before the first amplifier to attenuate interference from the 7.2075 GHz reference tone used during observations. This pre-amplifier positioning reduces the amplification of the interfering signal, thereby preserving the clarity and fidelity of the astronomical data \citep{gbt1}, albeit at the cost of added system temperature and decreased sensitivity at nearby frequencies.

In addition to hardware-based mitigation, the GBT leverages advanced tools to provide observers with real-time information on the RFI environment. These tools include a graphical user interface (GUI) and an interactive web-based platform that displays RFI scans, allowing observers to adjust their observational strategies dynamically. The combination of switchable notch filters and proactive monitoring tools significantly enhances the telescope’s ability to manage RFI, ensuring higher-quality observations and more reliable data.

By integrating front-end notch filters, switchable configurations, and complementary real-time monitoring systems, these observatories demonstrate the critical role of adaptive and innovative RFI mitigation strategies in maintaining the integrity of astronomical observations.

In summary, while front-end notch filters can offer targeted mitigation against severe RFI, they are deployed sparingly due to their detrimental effect on sensitivity. Their use underscores the trade-offs that observatories must navigate between interference protection and scientific performance.

\subsubsection{Pre-LNA superconducting filters}
The current RFI mitigation framework at the Yebes Observatory, in Spain, utilizes high-temperature superconducting (HTS) filters positioned ahead of the low-noise amplifiers. While effective and essentially noise-free, these filters present inherent limitations, including resonances or spurious responses at other frequencies within the receiver’s band. Moreover, as permanent components of the system, they cannot be automated. Recent advancements in filter design, such as the superconducting spiral bandpass filter developed by \citep{huang2018superconducting} using a pseudo-Fourier technique, have demonstrated enhanced interference suppression capabilities. The optimization of cryogenic receivers using superconducting filters has been further investigated in \citep{garcia2023optimizacion,lopez2021tri}, addressing both RFI suppression and reference signal phase measurement. Additional improvements are anticipated, with new filter designs currently under development and expected for publication.

\subsubsection{Digital: Excision in time domain}

The excision class of techniques are the most widely used RFI mitigation approach in the digital domain. An overview of the excision techniques and their broad classification is provided in \citep{buch2019rfic}.

\begin{itemize}
\item Real-time RFI Excision at uGMRT\\

The uGMRT operates from 120 to 1450 MHz and has near-seamless coverage in this band. It is an array of 30, 45m diameter parabolic dishes and a sensitive receiver which operates in the interferometry and beamformer modes. The array is located 80 km north of Pune in India. The increase in the population around the array in the last 30 years, along with the proliferation of communication devices, satellites, and mobile phones, has caused an increase in the RFI. The uGMRT signal processing backend has a real-time RFI excision system for mitigating RFI and achieving sensitivity levels close to the theoretical limits, defined by the radiometer equation.

Recording of the signal can be done, due to technical limitations, at lower time resolution only, especially in the case of interferometers. Therefore, filtering RFI signals in the signal chain at high time resolution allows for the removal of RFI, which would not be possible afterwards, such that part of the observation can be saved. Such a filtering cannot be applied in the post-recording domain, as this type of RFI spreads across the spectrum in the frequency domain. In general, the earlier the filtering is implemented, the less the loss of astronomical data.

\begin{figure}
    \centering
    \includegraphics[scale=0.7]{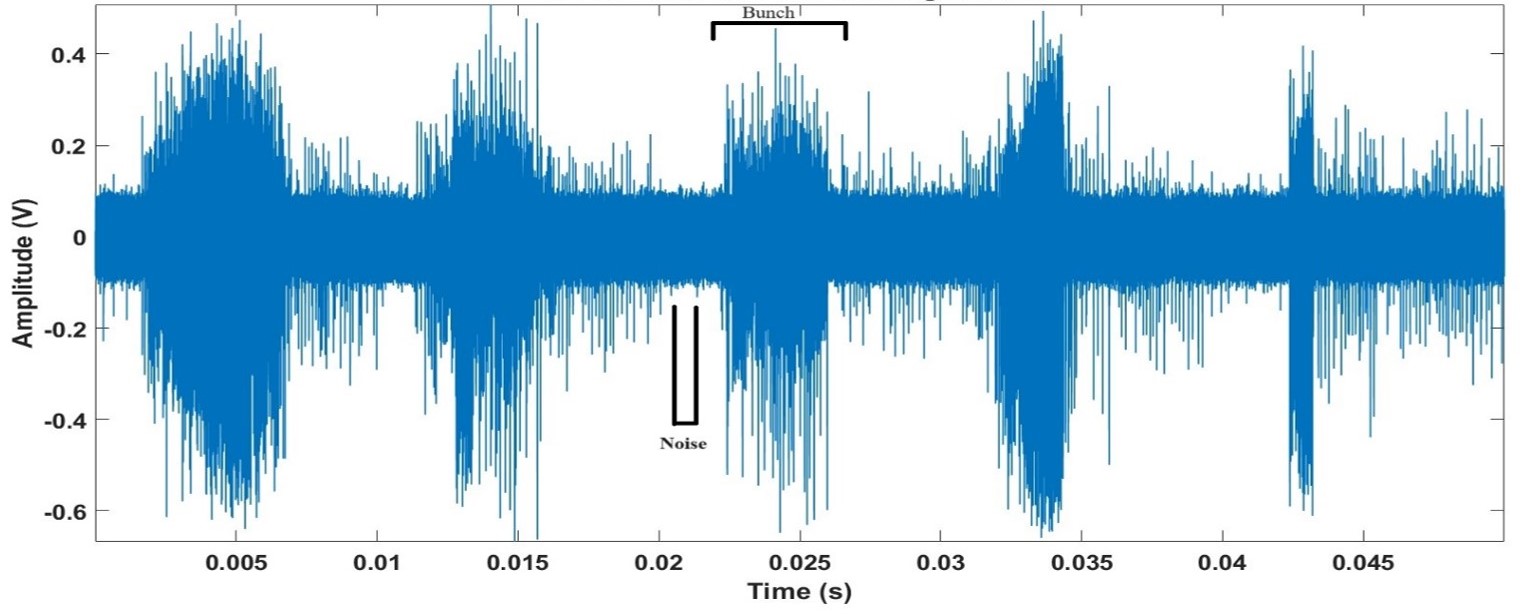}
    \caption{A 50ms time-series of single antenna Band-4 (550-850 MHz) uGMRT data. Powerline RFI can be seen to occur in bunches, repeating every 10ms. The data is acquired at the input of the signal processing system with a time resolution of 5ns.}
    \label{fig:ugmrt-b4-ts}
\end{figure}

One of the main causes of RFI at frequencies less than 1 GHz is sparking and corona discharge on the high-tension lines and transformer installations around the array. This type of RFI is impulsive in the time domain (Ref. Figure~\ref{fig:ugmrt-b4-ts}), resulting in a broadband increase in the spectral power. Since this type of RFI cannot be mitigated by frequency-selective filters in the receiver system, a real-time statistical RFI excision system was developed, which operates on digitized time series from each antenna and polarization. This system is designed to excise strong impulses in the received signal.

The real-time system currently implemented on Field Programmable Gate Array (FPGA) \citep{buch2019real} on ROACH-1 boards \citep{hickish2016decade} uses a Median Absolute Deviation (MAD) based robust estimation and threshold detection scheme. Each incoming sample (at Nyquist rate) is compared with a robust threshold, and samples detected as RFI are replaced with a constant value, threshold or digital noise \citep{buch2014variable} sample. After rigorous testing on various astronomical data products, the RFI system is released for observations and is currently extensively used for observations in the lower frequency bands of uGMRT ($<$ 1 GHz). An example of imaging through this system wherein half the antennas used the original signal, whereas the other half used the filtered copy \citep{buch2022performance} is shown in Figure~\ref{fig:ugmrt-b4-image}. 

\begin{figure}
    \includegraphics[scale=0.4]{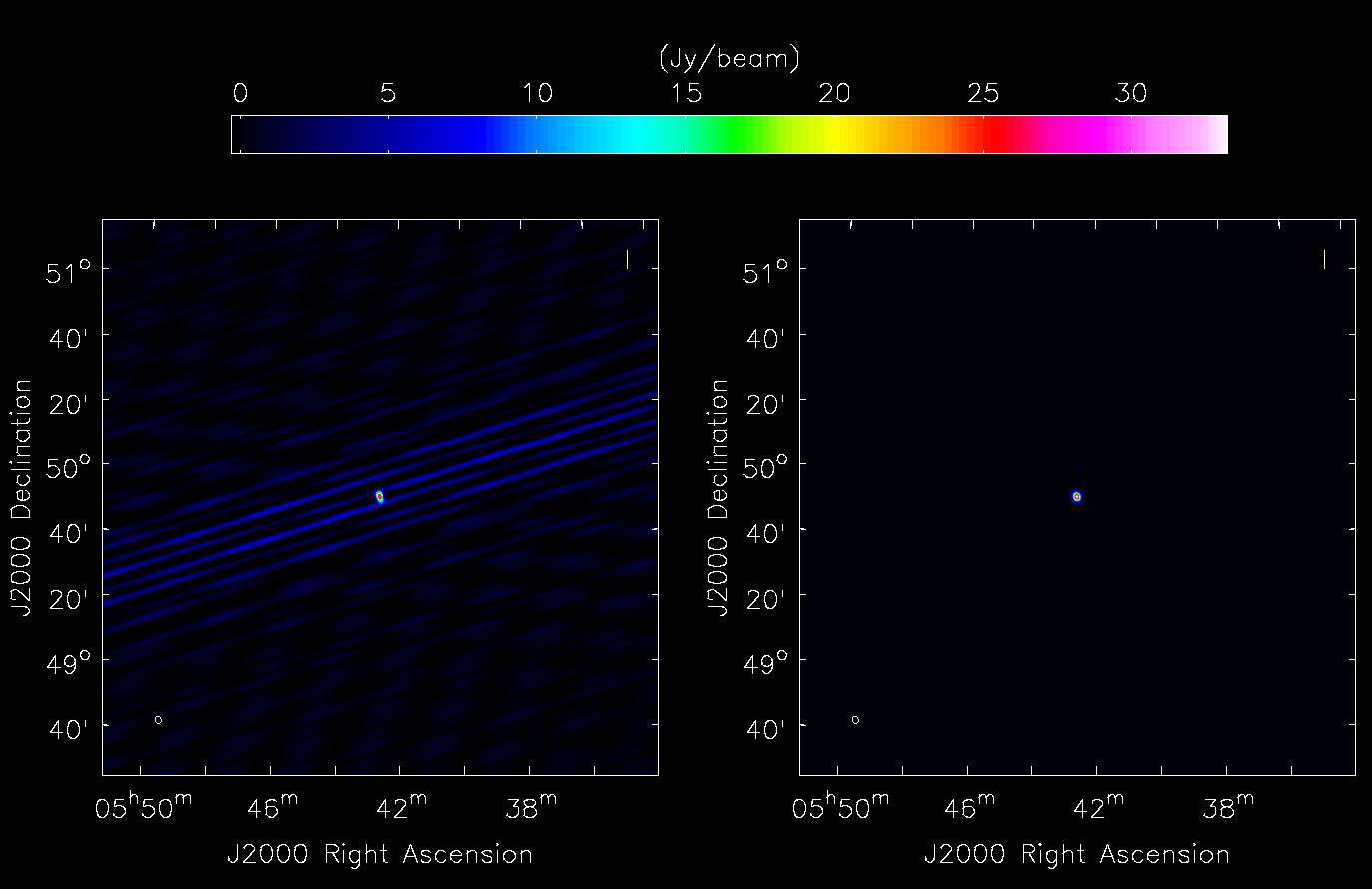}
    \caption{uGMRT image, 9 antenna, point source 3C48 observed in band-4 (550-850 MHz). The imaging is carried out using baselines $<$0.5km and tested simultaneously in the unfiltered and filtered modes to see the effects of real-time RFI filtering. The image corrupted by broadband RFI (left) is improved by a factor of 3 in the filtered version (right). The overall flagging is around 3\%. Image courtesy: Ruta Kale, NCRA, India.
}
    \label{fig:ugmrt-b4-image}
\end{figure}

A detailed report on the testing of the filter and effects on different types of astronomical data is provided in \citep{kale2025real, lal2025examining}. A possibility of refinements to the technique, including mitigating low-level RFI, using machine learning-based approaches, and implementing similar techniques in real-time for mitigating narrowband RFI \citep{buch2016towards, buch2016real} is being looked at. The technique used for uGMRT is being proposed \citep{buch2023real}for upcoming telescopes like the SKA. 

Some recently published scientific results show that the real-time RFI mitigation system was used for observing radio sources like the Fast Radio Bursts \citep{bethapudi2025rotation, bhardwaj2025constraining}, Radio Relics \citep{chatterjee2024new}, Radio Halo \citep{santra2024deep, manna2024radio}, Pulsars \citep{kumari2024first}, and the Sun \citep{mondal2025observation}.

\item Post-channelization flagger\\
The DSA-2000 is a next-generation radio telescope array designed to map the entire northern sky at high sensitivity, enabling unprecedented studies of transient phenomena and large-scale cosmic structure. Comprising 2,000 antennas spread across a large area, the DSA-2000 will provide continuous monitoring of the radio sky, particularly suited to detecting fast radio bursts and other dynamic cosmic events. The digital front-end of the telescope, referred to as the F-engine subsystem, manages key front-end processes, including signal conditioning through amplification and filtering, digitization, channelization, and packetization, enabling the transmission of raw data over a network to downstream processing nodes. As the only subsystem that receives the full processing bandwidth for each antenna, the F-engine is uniquely suited for identifying non-linearities that arise from harmonics or intermodulation products due to strong RFI. Detection of these non-linearities is achieved via two methods: high-cadence ($\approx$0.1 µs) monitoring of the total received power, compared to a calibrated operational range, and low-cadence ($\approx$0.1 s) assessment of spectral shape anomalies, based on root mean square differences between the integrated autocorrelation spectrum and a calibrated spectral model \citep{hellbourg2024flagging}. Both binary detection mechanisms will be implemented, with a logical OR operation to combine them, producing a flag to indicate non-linear responses. This flag is subsequently passed to the X-engine, responsible for the antenna signal correlations, where visibility data from affected antennas is either statistically adapted at lower integration times or fully nulled, depending on the impact of the non-linearity (RFI levels 2-3).
\end{itemize}

\subsubsection{Digital: Excision in spectral domain}
\begin{itemize}

\item Median-based Excision for uGMRT

A technique similar to that used at uGMRT for broadband RFI is proposed for mitigating spectral domain (narrowband) RFI. This technique will be implemented in the frequency domain for RFI from spatially confined sources. It operates in the frequency (post-correlation) and uses median-based spectrum estimation using the Median Absolute Deviation estimator. The technique takes care of equalising power across the band and ensures that RFI in the band where the power is low is also detected and excised effectively. The detailed algorithm is described in \cite{buch2016towards} with a tentative implementation plan in \cite{buch2016real}. Currently, this scheme is undergoing algorithmic refinements and real-time implementation to mitigate narrowband interference, particularly from spectrally and spatially confined sources.

\item Sigma Cut real-time flagger\\

RFIm \citep{sclocco2019real} is an open-source, high-performance library designed to mitigate RFI in real-time. RFIm is optimized to run on many-core accelerators such as GPUs and provides methods that are robust yet computationally efficient. The library features two main algorithms: Time-Domain Sigma Cut (TDSC) and Frequency-Domain Sigma Cut (FDSC). These two algorithms detect and replace RFI-contaminated data with statistical averages to reduce false positives without a significant impact on processing time.
While this replacement inevitably removes both RFI and any underlying astronomical signal, the algorithms are designed to act only on data that exceeds statistically defined thresholds. This ensures that most genuine astronomical signals, which typically follow expected noise distributions, are preserved. Furthermore, careful tuning of detection thresholds and the architecture of downstream pipelines helps recover transient events and minimize the loss of valuable data.

This trade-off allows systems like the Apertif Radio Transient System (ARTS) to maintain high data throughput while mitigating the impact of increasing anthropogenic and satellite-generated interference. However, some sensitivity to weak or rare signals may be sacrificed for the sake of real-time processing efficiency.
 
\item Higher-order statistics\\

The Spectral Kurtosis (SK) The technique suggested by \citep{nita2010statistics} is a statistical method used for real-time detection of non-Gaussian signals, RFI in most cases, within Gaussian noise in radio astronomy data. The SK estimator distinguishes signals by comparing power and power-squared values of spectral data to detect deviations from a Gaussian behaviour. The Expanded Owens Valley Solar Array (EOVSA) implements this SK technique in its correlator, using an FPGA-based system to perform real-time SK calculations directly within the F-engine. This design allows the system to flag and exclude contaminated data dynamically, enhancing the quality of observations by effectively identifying and mitigating RFI while preserving genuine astronomical signals.

\item MeFisTo: Frequency and Time Median filter\\

Between 5 and 88 MHz, the lowest part of the frequency band accessible to ground-based telescopes, the sky is crowded with audio and timing broadcast sources, allocated to frequency channels much closer than the frequency resolution required to meet the scientific needs of solar and Jovian observations. This band is also affected by wideband RFI generated by electric fences, spark ignition systems from internal combustion engines, and power lines.

Time and frequency analysis of this band must be performed with temporal and spectral resolutions exceeding the requirements for scientific observations. This results in data streams and data sets much larger than necessary for scientific purposes.

Since 2013, a receiver for the Nançay Decameter Array \citep{lecacheux2013, 2017pre8.conf..455L} based on 80-MSps ADCs and a Stratix IV FPGA has implemented a real-time streaming Blackman-Harris-windowed 64k-FFT, followed by classical (averaged) time-frequency analysis using a Welch-averaged periodogram. Simultaneously, the same periodogram is computed with a median filter (kernel size=64) applied first on the frequency axis, then on the time axis (kernel size=128), to mitigate narrowband and wideband RFI in observations. The frequency resolution for RFI mitigation is as low as 1.2 kHz, but it is raised to 78 kHz for scientific purposes. The time resolution for RFI mitigation is 0.8 ms, but is increased to 104 ms to provide a sustainable data rate for 24/7 monitoring of the Sun and Jupiter (see Figure~\ref{fig:rfi_MeFisTo} for an illustration of the data improvement).

\begin{figure}[ht]
    \centering
    \includegraphics[width=\textwidth]{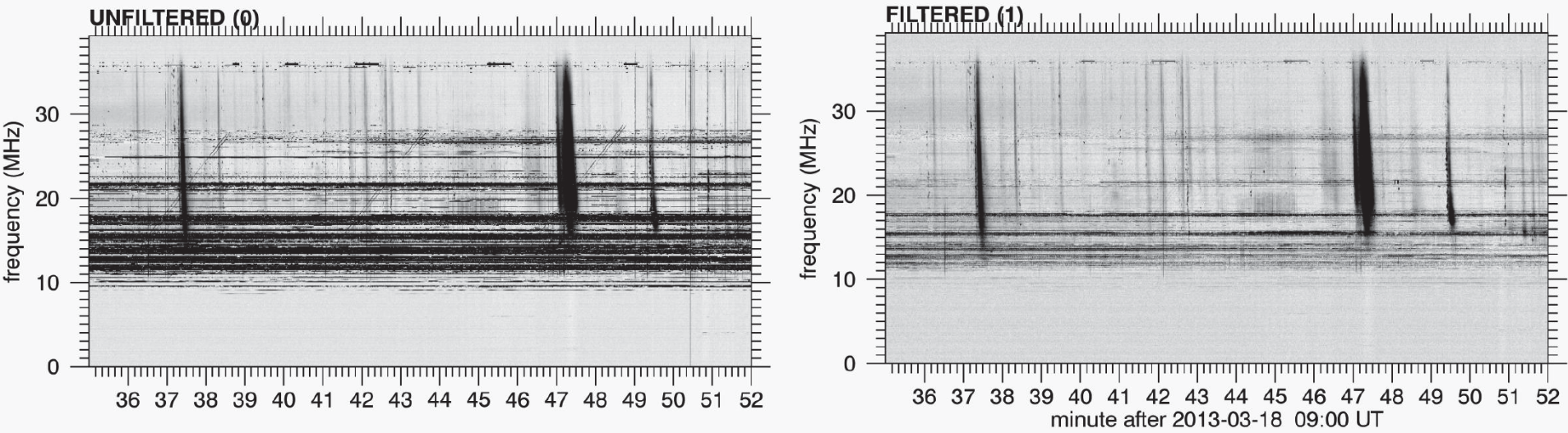}
    \caption{Solar activity observed without (left) and with (right) RFI median TF-filtering.  Wideband RFI and ionospheric sounders are eliminated, and shortwave broadcasts are significantly attenuated. \citep{lecacheux2013}}
    \label{fig:rfi_MeFisTo}
\end{figure}

\item Power-based excision RFI mitigation implementations\\

In the early 21st century, a configurable digital receiver (RDH: Reconquest of Hertz Domain) was deployed at the Nançay Observatory. The digital hardware, comprising ADCs (Analog to Digital Converters), FPGAs (Reconfigurable Logic Arrays) and DSPs (Digital Signal Processors), was shared by the 100-meter Decimeter Wavelength Telescope (NRT) and the Decameter Wavelength Analog Phased-Array (NDA). The main purpose of this receiver was to develop operational techniques for RFI mitigation. This hardware is now largely obsolete and was decommissioned years ago. However, the techniques remain and may be reused in future deployments.

The core work in designing these operational RFI mitigation techniques relies on a detailed study of the statistical behaviour of operators that can be easily implemented in real-time digital logic to create robust mean power estimators \citep{dumezviou:tel-00319939}. This toolbox can then be used to develop custom implementations that address the challenges encountered in the design of digital receivers for radio astronomy, such as:

\begin{itemize}
\item HI red-shifted radio galaxies have the rest frequency of the hyperfine transition at 1420 MHz shifted into frequency bands allocated to L-band radar systems used for civil and military aircraft detection and ranging. Every 5 or 10 MHz, a radar system emits several kilowatts of power for one microsecond, with a pulse repetition period of one millisecond. Theoretically, the sky is clear during the remaining 999 microseconds and available for astronomical observations. Detecting strong radar pulses with SNRs tens of dB above system noise is straightforward. The challenge lies in detecting weak pulses with negative-dB SNRs while maintaining an acceptable false alarm rate.

This was successfully implemented in the very limited resources of Virtex II FPGAs using the previously mentioned toolbox, allowing the excision of data blocks contaminated by both weak and strong radar pulses before they are passed to the spectral analysis block. As a result, HI surveys can now be conducted without artefacts originating from radar systems.

\item Red-shifted OH megamasers may have their 1665 and 1667 MHz lines fall into the band allocated to the Iridium satellite constellation. This system manages access to the radio spectrum through a Time-Frequency Division Multiple Access (TFDMA) scheme. During the test campaign, only a few per cent of the millisecond-scale instantaneous spectrum was occupied, but after a few seconds of integration, the entire band became corrupted.

Spectral analysis was configured to generate data with time and frequency resolutions (3.4 kHz and 2.34 ms) tailored to the TFDMA characteristics. A MAD-estimated power criterion was then applied to discriminate between corrupted and pristine data slots, removing RFI from the dataset before integration. Observation with and without this processing is presented in Figure~\ref{fig:rfi_Iridium_blanked}.

\begin{figure}[ht]
    \centering
    \includegraphics[width=\textwidth]{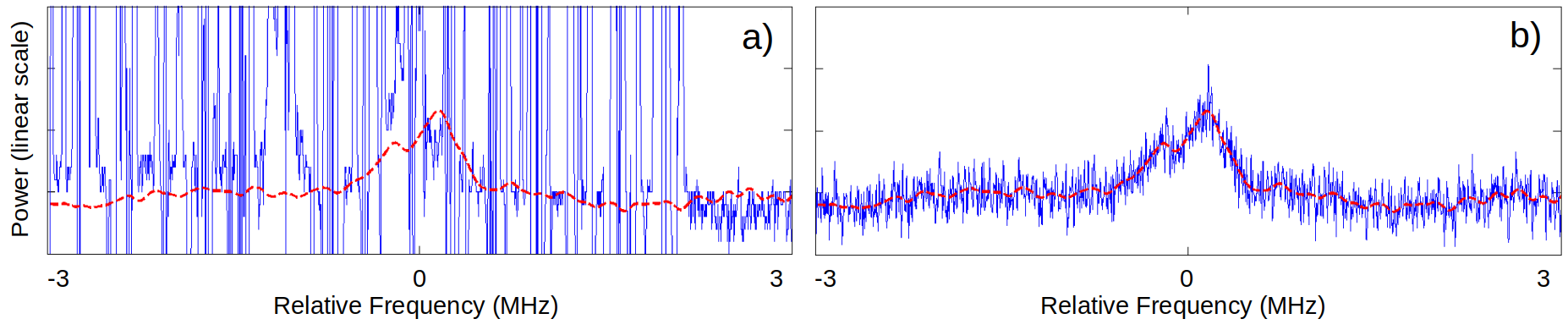}
    \caption{Re-observation of III Zw 35 on January 8, 2004, in real-time after 14 minutes of integration with the NRT: (a) Without "blanking." The vertical axis is adjusted to make the expected source profile visible. (b) With "blanking" (3.5\% of data eliminated). The source is visible again.}
    \label{fig:rfi_Iridium_blanked}
\end{figure}

This was implemented on TMS 6203 Digital Signal Processors chips from Texas Instruments to process the frequency band in real time.
\end{itemize}

\item Strong impulsive RFI may lead to false positives in the Canadian Hydrogen Intensity Mapping Experiment (CHIME) fast radio bursts (FRB) pipeline, and therefore needs to be mitigated \citep{chime_frb_rfi_2023}. Several real-time iterative processes have been applied to weed out the statistical outliers from the channelized intensity data that are effectively high-pass filtered.

The RFI mitigation process before the dedispersion transform in the L1-pipeline for CHIME/ FRB search is explained in \citep{chime_frb_rfi_2023}. Dedispersion is the process of reversing the frequency-dependent time delay introduced by the interstellar medium, aligning the arrival times of a dispersed radio pulse across frequency channels. It is essential for recovering the sharp time profile of fast transients such as FRBs.

The channelized intensity data of each beam are processed by a `sub pipeline' containing alternate \textit{Clipping transforms} and \textit{Detrending transforms chains}. There are two different clipping transforms: intensity and standard deviation clipping transforms, to mask the statistical outliers in the intensity data. Every clipping iteration helps
improve the RFI mask by recognizing more statistical outliers and hence reshaping the masked intensity Probability Density Function (PDF) to a robust $\chi^2$ distribution in real-time. The large-scale variations from RFI, forward gains, and digital beamforming are seen in the CHIME/FRB intensity time series as functions of time, frequency, and sky location. Due to the distorted intensity power distribution function, the clipping and dedispersion transforms fail. The proposed detrending transforms provide a computationally less-intensive way of high-pass filtering in the harmonic space of intensity.

\item Correlator Outlier Excision\\
The MeerKAT correlator's initial data ingest stage employs a streamlined, one-dimensional mean absolute deviation (MAD) filter, termed ``ingest\_rfi'', which operates along the frequency axis to independently threshold each data dump and correlation product. Implemented on a Graphics Processing Unit (GPU), this filter is conservatively tuned to detect prominent amplitude spikes, flagging approximately 4\% of the L-band data for imaging modes, excluding pulsar observations, which are subject to a specialized RFI mitigation approach. Originally, flagged data from ``ingest\_rfi'' was excised before averaging the 0.5-second raw correlator dumps into the final 8-second dumps, meaning that RFI-contaminated segments were removed before integration. However, this approach was ultimately revised based on feedback from the scientific and RFI research communities.
Currently, the 0.5-second data are averaged into 8-second dumps without excision based on ``ingest\_rfi'' flags. Instead, these flags are retained as metadata and used to inform downstream processing steps.
The ``ingest\_rfi'' flags are applied conservatively, while the more assertive ``cal\_rfi'' flags are typically excluded. Observers utilizing the MeerKAT pipeline images receive outputs with all standard flags applied. Further details can be found in the MeerKAT flagging package documentation \citep{hugo2022tricolouroptimizedsumthresholdflagger}.

\item Calibration VarThreshold\\
The calibration pipeline following the ingest stage at MeerKAT incorporates an advanced, two-dimensional RFI flagging algorithm, ``cal\_rfi'', custom-developed in Numba as an adaptation of the AOFlagger \cite{offringa2010lofarrfidetectionpipeline}. This algorithm identifies background noise across both time and frequency axes in two stages: an initial pass to establish a smooth baseline on unflagged data, followed by a sum-thresholding step \cite{sihlangu2019meerkat}. Flagging is selectively applied to HV and VH (cross-hand) polarizations, with the resulting flags extended to HH and VV data. During calibration, ``cal\_rfi'' flags are set before gain determination for the calibrator and reapplied after gain corrections on the target, enabling more stringent RFI rejection parameters. This procedure typically flags about 20\% of L-band data across all imaging modes, with pulsar modes exempted, as they employ a distinct RFI management strategy.

\end{itemize}

\subsection{Prospective real-time implementations}
\label{subsection:hardware:catalog:prospective}


\subsubsection{Preventive: satellite avoidance at Green Bank}
\label{subsubsec:avoidance}
Satellite boresight avoidance is a technique to prevent satellites from directly illuminating a telescope's primary observation direction. In experiments conducted by the GBT with Starlink satellites, this technique was tested \citep{nhan2024spectrumcoexistencedemonstrationeffectiveness} by adjusting the satellite constellation's configuration based on real-time telescope pointing data and predicted satellite trajectories. This approach reduces interference by dynamically coordinating telescope operations with satellite activities, such as emission control and beam steering. The process is facilitated by the Operational Data Sharing (ODS) system, an autonomous platform developed by the National Radio Astronomy Observatory (NRAO). ODS provides satellite operators with real-time updates on telescope positions and observing frequencies, allowing them to adjust satellite operations accordingly. With frequent updates, potentially every minute, ODS ensures effective coordination, especially during close satellite passes near the telescope’s boresight. How effective the boresight avoidance technique is, also at different kinds of telescopes, is still subject to further research \citep{boresightavoidancereport}.

\subsubsection{Analog: Tunable notch filter DSA-2000}

The tunable notch filters are designed to mitigate RFI in the analog front end of the DSA-2000 receivers \citep{hellbourg2024assessing}. The Quad-Stub Resonator Filter uses four high-Q “series LC” stubs, each controlled by voltage-tuned varactors, allowing independent frequency tuning for precise RFI mitigation. The filter achieves deep notch attenuation, adjustable within a range of 550 MHz to 1050 MHz and 1600 MHz to 2 GHz, with peak attenuation up to 62.5 dB. It features a chassis design with RFI-tight seals and SMA connectors to prevent interference coupling. The filter's response is fine-tuned through individual varactor bias adjustments, enabling adaptable and effective RFI suppression across a broad frequency range.

While the filter is placed before the first LNA, its high-Q design ensures minimal insertion loss. Nevertheless, a modest increase in receiver temperature (\(T_{\text{rec}}\)) is expected, which is carefully characterized during system calibration. This trade-off is justified by the significant improvement in dynamic range and the ability to suppress strong, persistent RFI without saturating the front end.

\subsubsection{Analog: Reconfigurable Intelligent Surface}

Reconfigurable Intelligent Surfaces (RIS) can mitigate RFI at the telescope receiver by dynamically shaping the electromagnetic wavefronts to create a destructive interference zone around the receiver \citep{zou2022scisrs,wei2024ris,wei2023multistage}. The RIS array, composed of multiple controllable elements, adjusts the phase and amplitude of reflected signals to precisely counteract incoming RFI. By steering reflected signals from the RIS in such a way that they are out of phase with the incident RFI, the RIS effectively cancels the RFI energy before it reaches the telescope’s receiver. This approach allows for the creation of an electromagnetic quiet zone, enabling the telescope to perform sensitive astronomical observations without interference from external RFI sources, such as aircraft and satellites, without altering the telescope’s primary astronomical signals.


\subsubsection{Digital: cyclic spectroscopy Green Bank}

The GBT is actively developing a cyclic spectroscopy system aimed at high-resolution studies of pulsars and improving RFI mitigation \citep{dolch2021deconvolving}. Cyclic spectroscopy is particularly valuable for separating the periodic signals of pulsars from RFI, allowing the system to maintain detailed phase and frequency resolution. This technique enhances the GBT’s ability to study narrowband interstellar scintillation structures and scattering delays, which can be distorted by RFI. By accurately capturing the cyclic structure of pulsar signals, cyclic spectroscopy helps distinguish astronomical signals from interfering sources like satellite transmissions, which lack this periodicity. This discrimination is typically combined with rejection or filtering algorithms that suppress non-cyclic components, ensuring that only signals with the expected cyclostationary signature of pulsars are retained.

Moreover, cyclic spectroscopy at the GBT is complemented by other RFI mitigation techniques, such as spectral kurtosis, which monitors departures from Gaussian noise to identify RFI-contaminated data segments. This approach has proven effective across different astronomical sources observed at GBT, improving data quality by flagging and replacing RFI with representative noise, thus preserving the integrity of the scientific data collected.

\subsubsection{Digital: Excision in time parametric subtraction}

The time-domain coherent cancellation (CTC) technique is a method used to mitigate RFI by subtracting an interference estimate from the received signal in real-time \citep{ellingson2022coherent}. CTC operates by generating a reference signal that represents the interfering source, which can be acquired externally (e.g., through a separate antenna) or synthesized internally using prior knowledge of the interference. The reference signal is used to create an interference estimate that is coherently subtracted from the received signal, ideally leaving only the astronomical signal of interest with minimal added noise or distortion. The approach allows telescopes to "look through" interference, preserving affected data that would otherwise be discarded by traditional methods
However, the technique has notable limitations. It requires precise estimation and synchronization, and any mismatch between the reference signal and the actual interference due to path differences, channel distortions, or nonlinearities leads to residual RFI that cannot be fully cancelled. Moreover, CTC cannot mitigate distortions introduced upstream in the receiver chain, such as non-linear effects or saturation in the analog front end. In such cases, even perfect subtraction of the interfering signal cannot recover the original astronomical signal, making prevention of saturation a prerequisite for effective cancellation.

\subsubsection{Digital: Threshold-based pre-correlation RFI detection}
\label{subsection:hardware:catalog:ska-mid}

\begin{figure}
    \centering
    \includegraphics[height=.28\textheight]{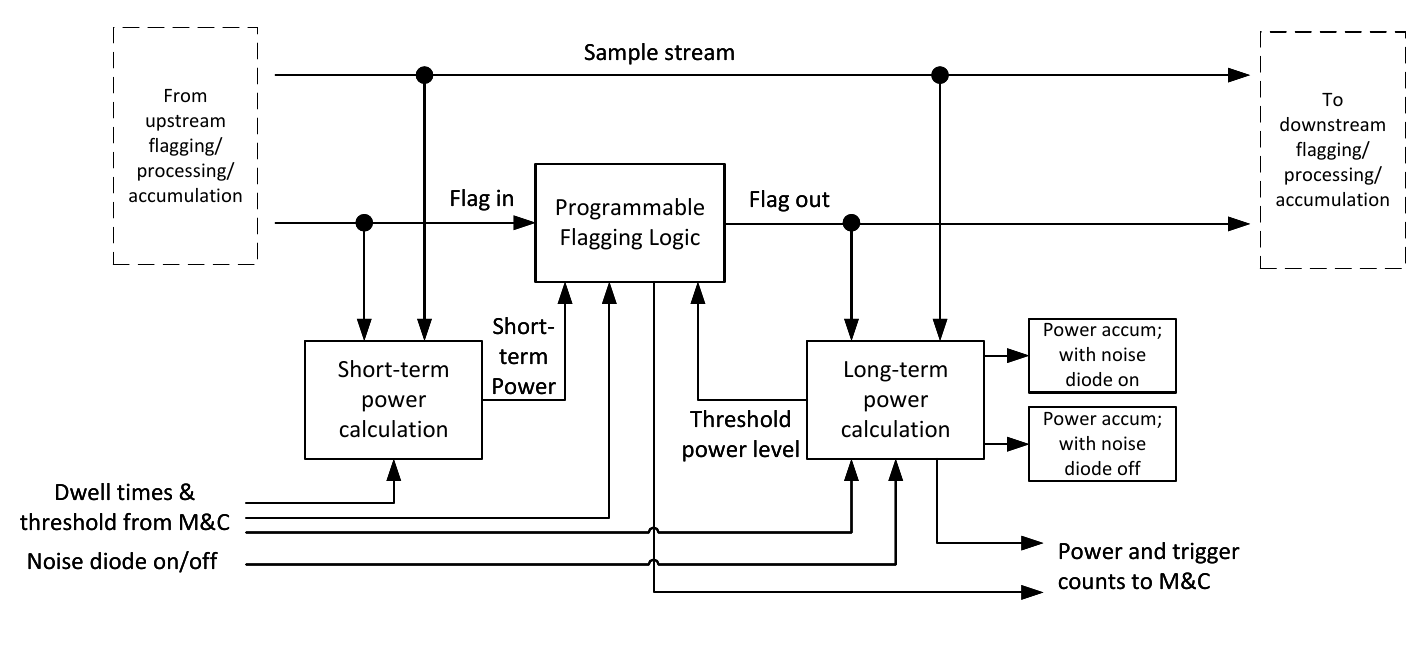}
    \caption{The proposed RFI threshold detector/flagger (from \citep{ska_mid_cbf_rfi_2019}).}
    \label{fig:rfi_df_ska_mid_cbf}
\end{figure}

The Correlator Beamformer for the Mid-frequency telescope for the \href{https://www.skao.int/en}{Square Kilometre Array} (SKA-Mid). CBF uses a `power threshold-based' RFI detection scheme \citep{ska_mid_cbf_rfi_2019}. The main objective is to preserve the linearity of the signal chain while achieving the desired dynamic range to maintain the sensitivity. The architecture of the proposed RFI detector/flagger module is shown in Figure~\ref{fig:rfi_df_ska_mid_cbf}. As shown there, two time scales are involved:
\begin{enumerate}
    \item Long-term ($>$1 sec, considering on/off state of the noise diode) for establishing an average power level, and
    \item Short-term to detect and flag bursting RFI to prevent downstream spectral splatter.
\end{enumerate}

The number of samples used for the short-term power calculation is programmable. If the short-term power calculation is greater than a programmable threshold,
then a flag is set and carried with the data stream to the downstream processing module, and downstream operations (i.e. any that affect output data products or intermediate calculations needed, such as gains/levels settings) are inhibited for the number of samples that it takes the flag to propagate through all downstream processing modules. As per the flagging policy, both polarization components are flagged if one is flagged. The dwell time of the flag is also programmable to balance the data loss with the risk of contamination. Note that the flagged data is not used in evaluating the long-term power calculation to have an estimate of the signal level of the non-RFI-contaminated signal. The long-term power is accumulated in two separate bins, one for the noise diode on and one for the noise diode off.

\subsubsection{Digital: excision in UV domain}

GRIDflag \citep{10464448} is a UV-plane-based RFI flagging algorithm designed to enhance the sensitivity and imaging fidelity of radio interferometric observations in the presence of strong and persistent RFI. Unlike traditional methods that flag RFI in the time-channel plane of baselines, GRIDflag operates directly in the UV plane, where multiple baselines sample similar spatial Fourier components redundantly across a regular grid. By statistically comparing UV samples within each grid pixel, GRIDflag identifies and flags RFI-affected points, preserving UV coverage and sensitivity to spatial scales.

The concept of UV-plane-based flagging has been explored previously and is implemented in pipelines like CARACal for offline use \citep{jozsa2020caracal}. However, GRIDflag introduces a novel real-time implementation of this concept, tailored for interferometers such as the DSA-2000. In these architectures, functioning as 'radio cameras' with 2000 elements, visibility data are not stored due to bandwidth and storage constraints. Instead, visibilities are computed and processed on-the-fly, making real-time UV-plane RFI mitigation essential.

This approach is particularly effective in mitigating systematic noise introduced by faint RFI, which can otherwise degrade image sensitivity and accuracy. GRIDflag's design allows it to be computationally efficient and adaptable to modern technologies like GPUs, making it well-suited for current and next-generation radio telescopes.

\subsubsection{Digital: Spatial filtering}
\begin{itemize}
\item Classic spatial filtering

The core idea is to leverage the spatial diversity of RFI and astronomical signals, along with the unique spatial signatures of interfering sources, to separate and suppress RFI \citep{hellbourg2014radio,hellbourg2016spatial,hellbourg2014rfi}. By estimating the RFI subspace using techniques like Singular Value Decomposition (SVD) and exploiting cyclostationary properties of RFI signals, spatial filters are designed to project received signals onto subspaces orthogonal to the RFI, effectively nulling interference. These filters can be applied either before or after correlation stages in the data processing chain. The technique is particularly valuable for modern radio telescopes, such as Low Frequency Array (LOFAR) and Electronic Multibeam Radio Astronomy Concept (EMBRACE), where dense antenna arrays provide numerous redundant measurements, enhancing the precision of RFI mitigation without significantly impacting the desired astronomical signal. This approach helps maintain high fidelity and sensitivity in observations despite increasing RFI challenges from human-made sources.

\item Reference-antenna

A reference antenna can be used to mitigate RFI in real time by providing an accurate estimation of the RFI spatial signature, which can then be used for spatial filtering in radio astronomy. The reference antenna is positioned to specifically capture the RFI without significantly receiving the astronomical signals. This RFI reference signal is correlated with the signals received by the main telescope array, allowing the RFI's spatial signature to be extracted and identified. This information is then used to project out the RFI from the main array's data using spatial filtering techniques, such as subspace projection.

By continuously updating the RFI spatial signature using subspace tracking algorithms, like the Power Method or Rayleigh Quotient Iteration, the system can adapt to changes in the RFI environment in real time, effectively cancelling out the interference without impacting the astronomical signals. This approach is particularly effective for managing multiple RFI sources and dynamically changing interference patterns, enabling high-fidelity observations even in heavily contaminated radio bands\citep{hellbourg2014reference,sardarabadi2015spatial}.

\end{itemize}
\subsubsection{Digital Artificial Intelligence (AI)/ Machine Learning (ML)}
\begin{itemize}
\item Collaborative Signal Subtraction

The collaborative approach to RFI mitigation works by establishing a bidirectional communication between radio telescopes and interfering sources, such as cellular networks, to actively share RFI information and cancel interference in real time. The interfering sources decompose their signals into a compact representation using techniques like the Karhunen–Loève Transform (KLT), which captures the signal's statistical properties independent of time and frequency. This RFI information, including eigenfunctions representing the interference, is periodically shared with the telescope through a low-overhead communication channel.

At the telescope, the received composite signal, which includes both astronomical signals and RFI, is similarly decomposed. The shared RFI characteristics are then used to cancel the interference from the telescope’s data through orthogonal projections in the eigenspace, thereby revealing clean astronomical signals. This approach significantly reduces communication and computational overhead by prioritizing the sharing of high-impact RFI data and selectively updating the interference model based on its influence on the telescope, thus maintaining high accuracy in signal recovery while managing multiple RFI sources effectively \citep{chakraborty2023collaboration,chakraborty2024low}.

However, the technique has several limitations. It requires active participation and coordination with interfering emitters, which may not be feasible in all regulatory or commercial contexts. Additionally, the method relies on the assumption that the interference is sufficiently well-modelled by the shared eigen-decomposition: highly non-stationary, bursty, or nonlinear emissions may violate this assumption and lead to incomplete cancellation. Finally, any mismatch between the transmitted and received interference models (due to propagation effects, timing offsets, or dynamic fading) may introduce residuals that contaminate the astronomical data. These factors constrain the generalizability of the approach and currently limit its deployment to cooperative scenarios with known and well-characterized RFI sources.

\item Pattern recognition

Automated pattern recognition to identify RFI in real-time radio astronomy data \citep{Winkel_2007} uses edge detection and window fitting to isolate high-intensity regions in the time-frequency domain, likely corresponding to RFI. Statistical analysis then helps differentiate RFI from genuine astronomical signals by comparing intensity patterns and variations. Iterative baseline fitting further refines the detection to ensure that RFI is accurately identified without impacting the integrity of the astronomical data. This method can be implemented in real-time systems, enabling dynamic flagging and mitigation of RFI, which is crucial for maintaining data quality in environments prone to interference, such as those near populated areas or satellite systems.

\item External flag generation

An RFI detector can provide real-time flags for radio telescopes by using ML techniques to automatically identify and classify interference in the RF spectrum. The system described here \citep{9111666} employs a deep neural network (DNN) that processes time-frequency data, detecting and drawing bounding boxes around RFI events. This approach allows the detector to continuously monitor the RF environment and flag novel or anomalous RFI events in real time. The detected RFI events are flagged based on their spatial, spectral, and temporal characteristics, allowing the telescope to dynamically adjust its data processing pipeline to exclude or compensate for contaminated signals.

By integrating with the telescope’s data stream, the system can provide immediate notifications about RFI, enabling telescopes to respond promptly to interference without manual intervention. This real-time detection capability is particularly useful for maintaining the quality of astronomical observations by preventing the degradation of data due to unwanted interference.



\end{itemize}

\subsubsection{Sample Frequency Offset Sampling}

The pioneering Sample Frequency Offset Sampling method offers a systematic approach to mitigate the impact of strong out-of-band RFI contaminating the observed spectrum \citep{carlson_scfo_2017}. The Correlator Beamformer of SKA-Mid.CBF is the first radio telescope component to implement this method \citep{ska_mid_cbf_rfi_2019}. This method can also reduce the impact of artefacts generated in connection with the sample clock signals. 

In the proposed method for Sample Clock Frequency Offset (SCFO) sampling, each receptor in SKA-Mid.CBF is sampled at different sample rates, causing the strong RFI to leak into different Frequency Slices (FSs) at different frequencies. As a result, when cross-correlated, the leaked RFI components decorrelate based on the integration time and frequency offset. Additionally, the 'Shift-Resample-Shift-Back' method is used to mitigate the impact of strong RFI with high time occupancy that is concentrated in specific frequency bands (e.g., GSM, Digital TV, and GNSS). This method leverages the Frequency Slice Architecture in Mid.CBF, where each FS processes an approximately 200 MHz portion of the observing band. The FSs are shifted to position the strong RFI near the 0 Hz mark and then resampled to correct for SCFO sampling and delay-tracking, before being shifted back to their original frequency sense. While resampling with a 'fractional-delay filter-bank' introduces systematic sample-time jitters that create replicas of the input spectrum, adding a random delay-jitter can suppress these replicas at the cost of increasing the system's noise floor. However, the added noise power is directly related to the magnitude of the resampling error. By relocating the strong RFI close to the 0 Hz mark, the absolute error due to resampling is minimized, resulting in a lower system noise floor.

\section{Summary table} 
\label{section:hardware:summary}

A summary of real-time RFI mitigation techniques, along with the category, location in the system, and maturity, is provided in Table~\ref{real-time-tech}.

\begin{table*}[htbp]
  \begin{center}
  \caption{Summary of real-time RFI mitigation techniques. The maturity level distinguishes conceptual (i.e. theoretical), prototype (i.e. not yet deployed or under commissioning), implemented (i.e. deployed but optional), and in-use (i.e. in operation) techniques. The F-engine includes techniques in the time domain, and those implemented post-channelization and pre-correlation. The last column summarizes known benefits or limitations when quantified.}
  \label{real-time-tech}
  \scriptsize
  \begin{tabular}{|l|l|c|c|p{5.7cm}|} \hline 
\textbf{Category} & \textbf{Technique} & \textbf{System Level} & \textbf{Maturity} & \textbf{Benefits / Limitations} \\ 
\hline
Avoidance & Automated Scheduling & Scheduler & Prototype & Avoids high-RFI windows; depends on accurate RFI forecasting \\ \hline
Avoidance & Dynamic Scheduling & Scheduler & Conceptual & Flexible, but requires real-time RFI maps and coordination \\ \hline
Avoidance & Satellite Boresight Avoidance & Scheduler & Prototype & Can mitigate >20dB from LEOs; complex implementation \\ \hline
Analog & Adaptive Analog Attenuators & Analog receiver & In use & Prevents saturation but raises $T_{\text{sys}}$ during daytime by up to 30–50\%  \\ \hline
Analog & Pre-LNA Superconducting Filters & Analog receiver & Prototype & Raises receiver noise temperature by $<1$–$2$K; selective and narrowband \\ \hline
Analog & Front-End Notch Filters & Analog receiver & In use & Increase $T_{\text{sys}}$ by $\sim$3–8K depending on position and bandwidth  \\ \hline
Analog & Switchable Notch Filters & Analog front-end & In use & GBT: reduces 7.2GHz tone contamination; increases insertion loss marginally \\ \hline
Analog & Tunable Notch Filters & Analog front-end & Prototype & Up to 62.5dB suppression; adds ~3K to $T_{\text{rec}}$ due to filter insertion  \\ \hline
Analog & Reconfigurable Intelligent Surfaces (RIS) & External & Conceptual & Potential 15–30dB spatial suppression; experimental \\ \hline
Digital & Real-Time RFI Excision & F-engine & In use & Preserves uncorrupted data; $\approx$4–6\% data flagged (e.g., MeerKAT)  \\ \hline
Digital & Post-Channelization Flagger & F-engine & Implemented & Enables frequency-specific flagging; computationally light \\ \hline
Digital & Sigma Cut Flagging & Beamformer & Implemented & Effective real-time suppression; signal replacement reduces transients detection  \\ \hline
Digital & Spectral Kurtosis (SK) & F-engine & Implemented & Captures non-Gaussianity; may misclassify weak bursts \\ \hline
Digital & MeFisTo Filtering & Beamformer & In use & Up to 20–30dB filtering in SKA pathfinders (MeerKAT) \\ \hline
Digital & Power-Based Excision & Beamformer & In use & Cuts strong impulsive RFI; misses continuous weak sources \\ \hline
Digital & Calibration VarThreshold & Correlator & In use & Avoids skewing calibration solutions; operates at $\sim$8s resolution \\ \hline
Digital & GRIDflag (UV-Domain) & Correlator & Prototype & Reduces faint RFI contamination; not real-time unless in correlator pipeline  \\ \hline
Digital & Spatial Filtering & Beamformer / F-engine & Prototype & Effective in known source directions; residuals remain in sidelobes \\ \hline
Digital & Collaborative Signal Subtraction & F-engine & Conceptual & Enables signal salvation; residuals remain due to mismatch  \\ \hline
Digital & Pattern Recognition & Correlator & Conceptual & Can identify complex RFI; requires ML training and compute \\ \hline
Digital & Real-Time Satellite Prediction & Scheduler & Prototype & Avoids known satellite tracks; requires accurate TLE updates \\ \hline
Digital & Cyclic Spectroscopy & F-engine & Implemented & Discriminates periodic pulsar signals from non-cyclic RFI; may be combined with rejection filtering  \\ \hline
Digital & Sample Freq. Offset Sampling & F-engine & Prototype & Decorrelates RFI across stations; residual artifacts minimized by shifting RFI to 0Hz  \\ \hline
  \end{tabular}
  \end{center}
\end{table*}

\section{Conclusion}
\label{section:hardware:conclusion}

The future of radio astronomy faces significant challenges as the radio spectrum becomes increasingly congested with anthropogenic signals. The evolving landscape of RFI demands that observatories implement a variety of mitigation strategies, from real-time excision techniques to preventive measures like dynamic scheduling and collaborative frameworks with spectrum users.
The chapter provided an overview of the effects of RFI on radio telescope receivers, the classification of RFI, the need for real-time mitigation, and the contemporary real-time RFI mitigation techniques. As part of the real-time implementation, we covered analog and digital techniques in the radio telescope receiver systems. The chapter provides details on the contemporary techniques commissioned and proposed in various telescopes across the world. Many techniques, particularly for mitigating strong RFI in the analog domain, have yielded good results. Techniques in downstream digital signal processing have also shown benefits, particularly the Excision class. These are mostly the ones commissioned and used at the radio telescopes. More advanced signal processing techniques and recent ones using AI and ML are being explored. A summary table is presented at the end of the chapter.

Looking forward, integrating adaptive RFI mitigation into telescope design, alongside preventive measures like satellite boresight avoidance and reconfigurable intelligent surfaces, will be essential. As we push the boundaries of sensitivity in radio astronomy, robust RFI management will contribute to the protection of the integrity of observations and ensure the sustainability of astronomical research in an increasingly crowded spectrum.

\section{Acknowledgment}
\label{section:hardware:acknowledgment}
The authors thank various researchers based in observatories across the world for providing details of the contemporary and upcoming RFI mitigation techniques. The authors also thank their team members and colleagues for the observations and data presented in this chapter, and the IAU General Assembly 2024 Focus Meeting 5 organisers for inviting them to write this review chapter.
Some of the material used in this chapter is based upon work supported by the National Science Foundation under Grants Numbers 2229428, 2229497, and 2128497.

\bibliography{hardware}

\end{document}